# Exactly solvable model of sliding in metallic glass

Nikolai Lazarev, Alexander Bakai

*NSC Kharkov Institute of Physics and Technology, Kharkov, Ukraine*
n.lazarev@kipt.kharkov.ua

At low temperature, $T \to 0$, the yield stress of a perfect crystal is equal to its so called theoretical strength. The yield stress of non-perfect crystals is controlled by the stress threshold of dislocation mobility. A non-crystalline solid has neither an ideal structure nor gliding dislocations. Its yield stress, i.e. the stress at which the macroscopic inelastic deformation starts, depends on distribution of local, attributed to each atomic site, critical stresses at which the local inelastic deformation occurs. We describe exactly solvable model of planar layer strength and sliding with an arbitrary homogeneous distribution of local critical stresses. The macroscopic stress threshold of the athermal sliding is found. Kinetics of thermally-activated creep of the sliding layer is described. The rate of the thermally activated sliding is tightly connected with parameters of the low temperature strength. The sliding activation volume scales with the applied external stress as $\sim \sigma_e^{-\beta}$, where $\beta < 1$.

The proposed model accounts for mechanisms and the yield stress of the low temperature deformation of polycluster metallic glasses, since intercluster boundaries of a polycluster metallic glass are natural sliding layers of the described type.

**Keywords:** Yield strength; Athermal sliding; Creep; Activation volume; Polycluster

## 1. Introduction

The yield stress of a crystalline solid, $\sigma_Y^{cr}$, was first treated by Frenkel [1] who approximated the periodic force that is required to shear a perfect crystal by a sinusoidal function. He found that the maximum available stress, what is termed theoretical shear strength, is $\sigma_{th} = \mu_{cr}/2\pi$ where $\mu_{cr}$ is the shear modulus of the crystal. At $\sigma > \sigma_{th}$ cohesion between atomic layers breaks down. Later an improved estimations of $\sigma_{th}$ were obtained, see e.g. [2]. It appears that the reasonable estimate for $\sigma_{th}$ is given by

$$\sigma_{th} \approx 10^{-1} \mu_{cr} \tag{1}$$

It turned out that values of $\sigma_Y^{cr}$ measured experimentally are close to the theoretical strength (1) only for the defect-free crystals, e.g. whiskers. The yield stress of real crystalline materials containing dislocations is determined by the stress threshold of dislocation mobility. This parameter was first estimated by Peierls [3]. For metals it reads

$$\sigma_P \approx 10^{-4} - 10^{-3} \mu_{cr} \tag{2}$$

Equations (1) and (2) give the characteristic range of local critical stresses of a single crystal. The low temperature yield stress of a metallic glass (MG) has been found to be also proportional to the macroscopic shear modulus [4], but the proportionality factor differs from those in (1) and (2)

$$\sigma_Y^{MG} \approx 10^{-2} \mu_{MG} \tag{3}$$

While in a crystal the macroscopic value of $\mu$ is equal to its microscopic value [5], the shear modulus in MG is a random quantity. Its local value $\mu_i$ depends on the atomic configuration at site $i$. Thus $\mu_{MG}$ is a mean value. Usually the macroscopic shear modulus of an amorphous alloy is typically up to 30% lower than the shear modulus of the crystal of the same composition [4].

MGs possess noncrystalline disordered structure which can be characterized by a random potential relief with randomly distributed atoms in the potential minima [6]. To find the yield stress of MG one has to consider the problem of strength and thermally activated inelastic rearrangements of atoms within a shear layer where each atomic site is characterized by a random local critical stress which is needed for the inelastic rearrangements.

Different models of the inelastic shear strain in MG were developed. The first model of this type belongs to Argon [7-9]. According to Argon at low temperatures (below $0.8 T_g$, where $T_g$ is the glass transition temperature) MGs are deforming due to inelastic shear rearrangements of atomic groups composed of about 10 atoms. Later these carriers of plastic deformation were termed shear transformation zones (STZs) [10]. Taking into account that STZs are carriers of the plastic deformation, a set of equations of motion (roughly analogous to the Navier-Stokes equations for fluids) were deduced [10-13]. STZs velocity, density and orientation are dynamic variables. During deformation STZs appear and annihilate persistently. It has to be noted that in the Argon model MG contains STZs but its other structural properties are not specified. They are implicitly accounted for in the model parameters. To connect STZs with shear band formation it was assumed that the free volume [14-16] is properly redistributed and created in the deformed MG.

Later these ideas were utilized in similar approaches. For example, the cooperative shear model (CSM) [17] postulated that STZ involves ~$10^2$ atoms and that the shear transformation is a cooperative process similar to the α-relaxation in liquids. In computer simulations [18] the size of STZs was estimated to be about 1.5 nm. It should be noted that CSM is focused on initial stage of anelastic and inelastic shear transformations.

Important features of models based on the STZ concept are as follows. First, sliding is the main mode of the inhomogeneous inelastic and viscous-plastic deformation at low temperatures. Second, up to hundreds of atoms can be involved in STZ, and the linear size of an STZ is estimated to be of several nanometers or less. It means that peculiarities of MG structure on the scale of ~10 nm play a key role in the shear transformation and initiation of shear bands. Recent experimental investigations on tensile strength [19,20] revealed a strong size effect when the specimen size is less than 100 nm. These results indirectly point out the structure heterogeneities of about 10 nm in size.

The polycluster model of MG structure developed in [23-25] is based on the idea that the majority of atoms possess "perfect" non-crystalline local order. Groups of atoms with different locally preferred configurations (subclusters) are associated in non-crystalline clusters with narrow and stable intercluster boundaries. Validity of this model is confirmed by many direct and indirect experimental investigations. Direct examination of the MG structure by means of the high resolution field emission microscopy shows that both rapidly quenched amorphous ribbons and bulk MGs, obtained at relatively small cooling rate, possess a fine polycluster structure with characteristic sizes of clusters and subclusters about 10 nm and 1-3 nm respectively [21,22]. The boundary width is comparable with the atomic size. The mean binding energy of atoms within the boundary is lower than in the cluster body by the tenth of eV. Since the boundary density was found to be extremely high, ~$10^{-5} - 10^{-6}$ cm$^{-1}$, they play a dominant role in the plastic deformation of MG.

The mean value of the shear modulus of the cluster body is comparable to the shear modulus of a crystal and the strength of the cluster body is nearly equal to the theoretical strength. The local values of critical stress required for inelastic relocation of an atom within the boundary layer is a random quantity, being less than that in the cluster body. Therefore the intercluster boundaries are natural STZs. Shear transformations under applied stress appear first of all due to sliding within the boundary layers.

The problem of dislocationless sliding localized in a layer with a random distribution of local critical stresses was considered in [23-25]. Approximate solutions of equations derived there were obtained under the assumption that the distribution function of the local critical stresses is a spatially-homogeneous piecewise constant function. The obtained approximate solutions gave qualitative and, in some cases, reasonable quantitative description of the sliding process and allowed to predict the mechanism of the shear band formation on a qualitative level. Later this approach was applied for the description of MG hardening during partial crystallization [26].

In this paper we consider the problem of the sliding in a localized layer and strength of MG possessing the polycluster structure. The model [23-25] is developed further. The theoretical treatment is based on the exactly solvable model of the homogeneous sliding in layer with randomly distributed local strengths. This model is used for description of inelastic processes in MGs.

The paper is organized as follows. We start with formulation of the homogeneous sliding model in a disordered structure. Then kinetic equations of sliding in a planar layer with arbitrary distribution of local critical shear stresses are formulated. Rigorous solutions of the sliding master equations are obtained. It is shown that these solutions determine the macroscopic strength of sliding layer, sliding velocity and effective activation volume as functionals of the distribution function of the local critical stresses. Consideration of the heterogeneous plastic deformation of MGs is presented in the section 3. The last section contains a brief discussion and conclusions.

2. **Model of homogeneous sliding in disordered atomic layer**

In a non-metallic glass the elementary inelastic rearrangements of atomic configurations can be described as a) splitting-recombination, and b) translation of the broken bonds, Figs. 1a and 1b, correspondently. Rearrangements of the potential relief related to these inelastic deformations are shown in Figs. 1c, 1d. This picture is also applicable to MG although in this case the potential relief is much shallower, because metallic bonds are not as strong as covalent ones.

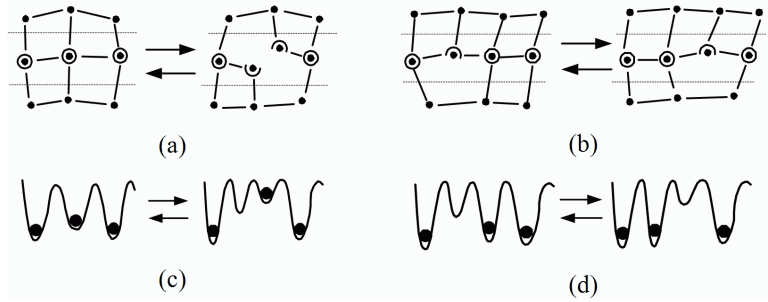

(a)                                    (b)

(c)                                    (d)

Fig. 1. Elementary rearrangements under shear stress: a) reversible splitting of a coincident site into two non-coincident sites (left to right) and recombination of two opposite non-coincident sites with subsequent formation of coincident site; b) transposition of coincident and non-coincident sites; c) and d) corresponding rearrangements of the potential relief. ● regular sites; ⊙ coincident sites; ☋ non-coincident sites.

In a polycluster the intercluster boundaries are regions of weak cohesion therefore here plastic deformation may occur by the sliding of one cluster over another. Figure 2a shows schematically the structure of sliding layer consisting of coincident and non-coincident sites. The potential relief in the sliding layer is shown in Fig. 2b. Due to strong alternation of sites of different types the formation of gliding boundary dislocations is mostly suppressed.

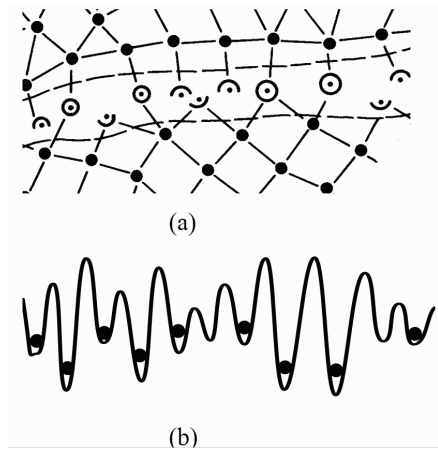

(a)

(b)

Fig. 2. (a) A fragment of sliding layer; (b) a schematic potential relief for sites in this layer.

## 2.1. Basic equations

When a planar layer is subject to an external shear stress $\sigma_e$, and atoms experience only elastic displacements then external stress is homogeneous in the layer. If some part of atoms underwent inelastic displacements this will cause stress redistribution i.e. stress relaxation at places of inelastic deformation and stress concentration at elastically deformed areas. Thus local stresses in sites are inhomogeneously distributed and time dependent. We assume that the local stress relaxation occurs by independent single-jump atomic rearrangements resulting in the external stress concentration at nondisplaced sites

$$\sigma_e^i = \frac{\sigma_e}{1-\Delta(t)} \qquad (4)$$

where $\Delta(t)$ is the fraction of displaced sites at time $t$.

The sliding velocity within the layer is a macroscopic quantity defined by the average frequency of inelastic displacements under the external stress

$$\upsilon_{sl} = <d>/\tau_{sl} \qquad (5)$$

where $<d>$ is the average site displacement during the elementary relocation, $\tau_{sl}$ is the average time for displacement of all sites of the slip layer per interatomic spacing.

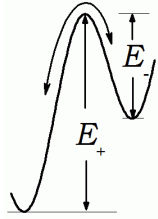

Fig. 3. Two-level potential for sites rearrangements.

The sliding time $\tau_{sl}$ is controlled by rates of individual inelastic displacements of atoms. Within a simple model of a particle in the two-level potential, see Fig. 3 and Appendix 1, the probability of thermally activated jump under external stress $\sigma_e$ can be expressed by

$$P(t,\sigma_i) = 1 - \exp(-\kappa(t,\sigma_i)), \qquad (6)$$

$$\kappa(t,\sigma_i) = \int_0^t \nu_0 \exp(-\alpha(\sigma_i - \sigma_e^i))d\tau$$

where $\nu_i^0$ is the vibration frequency in the site $i$, $\alpha = \upsilon_a/2k_BT$, $\upsilon_a$ is the atomic volume, $T$ is the temperature, $k_B$ is the Boltzmann constant, the parameter $\sigma_i = 2E_+/\upsilon_a$ is the critical shear stress when the atomic configuration goes from a site to a neighbor site without thermal activation. A physical meaning of parameter $\sigma_i$ is similar to the shear strength of a perfect crystal (1). Unlike to the crystalline lattice the amorphous solid is characterized by a wide spectrum of local critical stresses. Denote the distribution function of $\sigma_i$ by $g(\sigma_i)$.

The value $\Delta(t)$ has an evident relation to the probability of local inelastic displacements:

$$\Delta(t) = \int_0^\infty P(t,\sigma_i) g(\sigma_i) d\sigma_i \qquad (7)$$

Equation (7) is an integral equation with respect to $\Delta(t)$. The fraction $\Delta(t)$ enters the integrand $P(t,\sigma_i)$ via (4) and (6).

If relation

$$\sigma_i < \sigma_e /(1-\Delta(t)) \qquad (8)$$

holds true, site *i* has become displaced without thermal activation in a time as short as $\sim 1/\nu^0$. Therefore the integration in (7) is carried out not from zero, but from the lower limit

$$\sigma_0^* = \sigma_e/(1-\Delta_0(\sigma_e)), \quad \Delta_0(\sigma_e) = \int_0^{\sigma_e^*} g(\sigma_i) d\sigma_i \qquad (9)$$

Combining (7) and (9) we obtain the master equation

$$\Delta(t) = \Delta_0(\sigma_e) + \int_{\sigma_0^*}^\infty P(t,\sigma_i,\Delta(t)) g(\sigma_i) d\sigma_i \qquad (10)$$

## 2.2. Athermal sliding

As the first step let us consider Eq. (10) in the low temperature limit, $T \to 0$. In this case the second term in the RHS of Eq. (10) is equal to zero, and the athermal sliding starts when the macroscopic yield stress $\sigma_0^s$ is reached. Evidently $\sigma_0^s$ is a functional of the distribution function $g(\sigma)$. Because non-negativity of the distribution function $g(\sigma)$ the functional

$$\Delta_0(x) = \int_0^x g(\sigma) d\sigma \qquad (11)$$

is the single-valued function of $x$. Hence, the function

$$\sigma_e(x) = (1-\Delta_0(x)) \cdot x \qquad (12)$$

is also the single-valued function of $x$. Equations (11) and (12) determine a single-parametric relation between $\sigma_e$ and $\Delta_0$, thereby give the exact solution of the problem of athermal sliding.

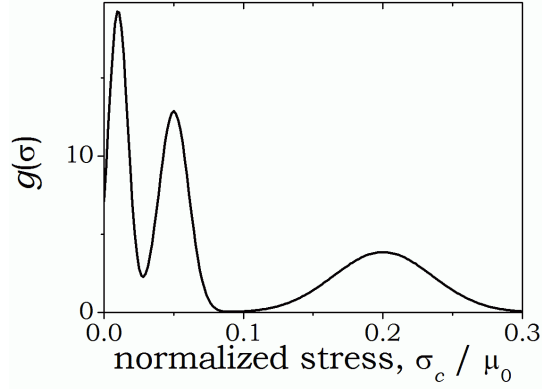

Fig. 4. Example of distribution function of critical shear stresses composed of 3 Gaussian distributions: $g(\sigma) = g_1(\sigma) + g_2(\sigma) + g_3(\sigma)$, with $g_k(\sigma) = g_k^0 \exp[-(\sigma - \sigma_k^0)^2 / 2\delta_k^2)]$.

As an illustration, the solution of Eqs. (11) and (12) is shown in Fig. 5 for the case of a trial distribution function $g(\sigma)$ depicted in Fig. 4. The points $\sigma^s$ are the solutions of the equation: $d\sigma_e/d\Delta_0 = 0$. At this points $\Delta(\sigma_e)$ changes in a stepwise manner with increasing $\sigma_e$. Using (9), (11) and (12) we obtain the equation for instability points $\sigma^s$ in the form:

$$x^s g(x^s) - \int_{x^s}^{\infty} g(\sigma) d\sigma = 0 \tag{13}$$

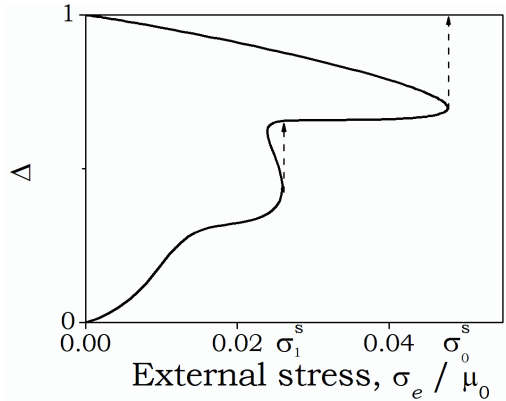

Fig. 5. Dependence of relative part of moved sites on external stress. Arrows show the changes in $\Delta(\sigma_e)$ when $\sigma_e$ increases.

Analysis shows that for any distribution function $g(\sigma)$ decreasing faster than $1/\sigma^2$ at $\sigma \to \infty$, equation (13) has at least one solution. The largest root of Eq. (13) defines the yield stress of athermal sliding $\sigma_0^s$.

Figure 5 shows that for the above-mentioned distribution function the threshold stress of athermal sliding $\sigma_0^s$ is approximately six times less than the maximal critical stress at the site.

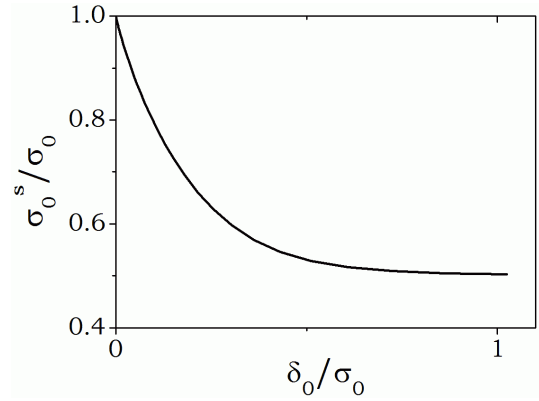

Fig. 6. Dependence of the athermal sliding threshold $\sigma_0^s$ on the variance of local critical shear stresses having truncated Gaussian distribution $g(\sigma) = g_0 \exp[-(\sigma-\sigma_0)^2/2\delta_0^2)]$, $\sigma \in [-\sigma_0, \sigma_0]$.

Because the variance of the critical stress values $\delta_0$ depends essentially on composition and thermal history of MG, it is important to investigate dependence of $\sigma_0^s$ on $\delta_0$. Taking $g(\sigma)$ in the form of truncated Gaussian distribution we have obtained the dependence shown in Fig. 6. One can see that $\sigma_0^s$ decreases fast when $\delta_0$ increases. At the variance $\delta_0 > \sigma_0/2$ the yield stress is nearly equal to $\sigma_0/2$, i.e. it is nearly 4 times less than the maximal value of critical stresses $\sigma_{max} = 2\sigma_0$.

### 2.3. Thermo-activated sliding

At low stress regime when $\sigma_e$ does not exceed the threshold of athermal sliding $\sigma_0^s$, Eq. (10) for the fraction of displaced sites is conveniently rewritten in the form

$$Z(t) = \int_{\sigma_e/Z_0}^{\infty} \exp(-\kappa(\sigma, Z)) g(\sigma) d\sigma, \quad \kappa(\sigma, Z) = \int_0^t \nu_0 \exp(-\alpha(\sigma - \sigma_e/Z(\tau))) d\tau \qquad (14)$$

where $Z(t) = 1 - \Delta(t)$. The parameter $Z_0$ is defined by relation (9).

The sliding time $\tau_{sl}$ can be found from the equation $Z(\tau_{sl}) = 0$.

To solve nonlinear integral equation (14), first, we transform it to the parametric form

$$Z(F) = \int_{\sigma_e/Z_0}^{\infty} \exp(-F\exp(-\alpha\sigma))g(\sigma)d\sigma \tag{15}$$

$$F(t) = \int_0^t v_0 \exp(\alpha\sigma_e/Z(F))d\tau$$

Differentiating $F(t)$ we obtain $dF/dt = v_0 \exp(\alpha\sigma_e/Z(F))$ or

$$\exp(-\alpha\sigma_e/Z(F))dF = v_0 dt \tag{16}$$

Integration of Eq. (16) gives the implicit solution for the function $F(t)$

$$v_0 \cdot t = \int_0^F \exp(-a\sigma_e/Z(F'))dF' \tag{17}$$

With $F \to \infty$, Eq. (17) specifies the sliding time of inelastic shear strain on interatomic spacing

$$\tau_{sl} = \frac{1}{v_o} \int_0^{\infty} \exp(-a\sigma_e/Z(F))dF \tag{18}$$

Here the functional $Z(F)$ is defined by Eq. (15). Equations (18) and (5) provide the analytical solution of the problem of thermo-activated slip in the disordered layer.

Figure 7 shows the results of numerical integrations of Eq. (18) for the distribution function $g(\sigma)$ depicted in Fig.4. The upper and lower solid lines show the dependence of strain rate on the external stress at low $T_1$ and high $T_2 = 2T_1$ temperatures respectively. Other curves correspond to intermediate temperatures. As it was expected, the curves for different temperatures meet at the point $\sigma_e = \sigma_0^s$, where the sliding time tends to zero.

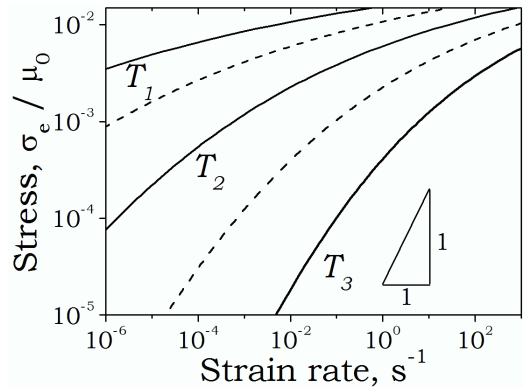

Fig. 7. The dependence of the strain rate on external stress at several temperatures: $T_1 \approx 0.5 \cdot T_g$, $T_2 \approx 0.7 \cdot T_g$ and $T_3 \approx T_g$.

The activation volume of sliding is not a constant; it depends on the external stress. Using definition [27]

$$V_{act} = k_B T \frac{\partial \ln(\tau_{sl})}{\partial \sigma_e} \tag{19}$$

we can calculate $V_{act}$ by application of Eqs. (18) and (15). Stress dependence of $V_{act}$ is shown in Fig. 8. Evidently the activation volume at low stresses is much larger than the atomic volume $v_a$ and scales with external stress as

$$V_{act}(\sigma_e) \sim \sigma_e^{-\beta} \tag{20}$$

The exponent $\beta$ depends weakly on temperature and lies in the range $0.5 < \beta < 1$. While stress is growing, $V_{act}(\sigma_e)$ decreases to the value which reaches the atomic volume when $\sigma_e$ approaches the yield stress $\sigma_0^s$.

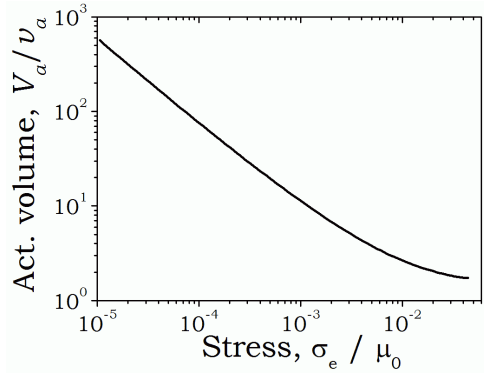

Fig. 8. Dependence of activation volume on the external stress.

Integration of Eq. (18) can be performed numerically, but in many cases it can be solved analytically. Making change of variable in Eq. (18), $F = \exp(\alpha x)$ we obtain

$$\tau_{sl} = \frac{\alpha}{v_0} \int_{-\infty}^{\infty} \exp(-\alpha \psi(x)) dx, \quad \psi(x) = \frac{\sigma_e}{Z(x)} - x \tag{21}$$

where according to (15) the function $Z(x)$ is given by

$$Z(x) = \int_{\sigma_e/Z_0}^{\infty} \exp(-\exp(\alpha(x-\sigma))) g(\sigma) d\sigma \tag{22}$$

Function $\psi(x)$ has a maximum at the point $x_0$ where $d\psi(x)/dx = 0$ and $d^2\psi(x)/dx^2 > 0$. The equation for $x_0$ takes the form

$$\frac{\sigma_e}{Z^2(x_0)}g(x_0)(1-\exp(-\exp(\alpha(x_0-\sigma_e/Z_o))))-1=0 \tag{23}$$

At low temperatures ($\alpha \gg 1$) this equation is simplified to

$$\frac{\sigma_e}{Z^2(x_0)}g(x_0)-1=0 \tag{24}$$

where $Z(x_0)$ is defined by (22).

The second derivative of $\psi(x)$ in the point $x_0$ is

$$\psi''(x_0)=2Z(x_0)/\sigma_e \tag{25}$$

Upon integrating (21) subject to (24) and (25) we obtain

$$\tau_{sl}=\frac{1}{v_0}\sqrt{\frac{2\pi\alpha\sigma_e}{Z(x_o)}}\,\exp(-\alpha\psi(x_0)) \tag{26}$$

Equation (26) was obtained under assuming that the distribution $g(\sigma)$ is a smoothly varying function in the vicinity of $x_o$. Using Eqs. (21) and (23) it can be shown that $\partial\psi(x_0)/\partial\sigma_e=1/Z(x_0)$. Therefore, it follows from Eq. (24) that $Z(x_o)$ depends on $\sigma_e$ as $\sim 1/\sqrt{g(x_0)\sigma_e}$. This result validates relation (20) for activation volume at low stresses.

### 3.  Heterogeneous deformation of MG

When considering the boundary sliding in polycrystals and polyclusters the finite sizes of boundary layers between triple joints of grains have to be taken into account. Boundary sliding is blocked in the triple joints. The sliding layer can propagate into the neighboring grain when the stress at the boundary edge of the triple joint exceeds a critical value $\sigma_c^*$, as it is shown in Fig. 9. Its edge is a one-dimensional dislocation-like defect which we call dislocation-like edge of sliding zone (DLESZ). In polycrystals the dislocation network and rotational motion of grains control the macroscopic plastic deformations. In polyclusters formation of a similar dislocation network is impossible because the dislocation sliding layer possesses the structure similar to that of intercluster boundary. Therefore a DLESZ is a carrier of inelastic deformation only if it is moving within the cluster body. When it reaches the cluster boundary, the cluster body becomes divided into two parts. In other words, DLESZ cuts non-crystalline grain into two parts.

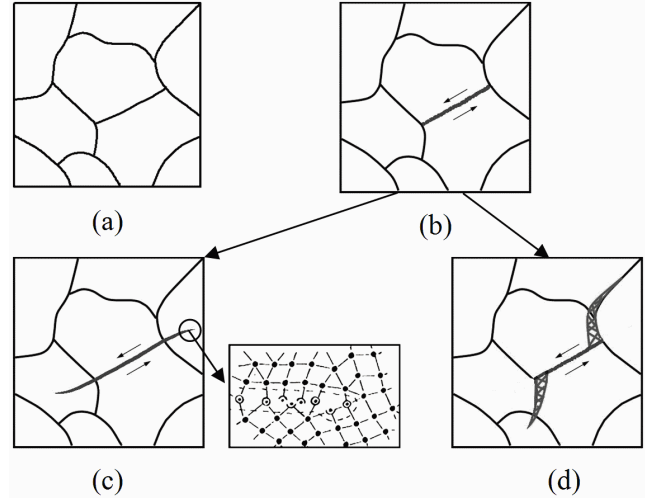

Fig. 9. Stages of the plastic deformation of a polycluster. a) Initial polycluster structure, the cluster boundaries are shown. b) Under external stress a DLESZ is formed as result of boundary layer sliding. The loop is blocked at triple joints of clusters. c) The DLESZ climbed into the neighboring clusters. In the insert the microscopic structure of the DLESZ is shown. Non-coincident sites are marked by semi-circles. (d) Nano-cracks along the intercluster boundary is initiated.

Initial stages of inelastic deformation of a polycluster are shown in Fig. 9. A fragment of the two-dimensional section is depicted in Fig. 9a. Boundaries demarcate clusters. In Fig. 9b initial deformation stage of the polycluster is shown at $\sigma_c^* > \sigma > \sigma_0^s$. Shear transformation occurs in the boundary section of size $l$. The shear stress is in plane of the boundary section. While the cluster bodies are elastically deformed, inelastic deformation takes place in the marked boundary section. It has to be noted that inelastic rearrangements take place at $\sigma < \sigma_0^s$. In this case the fraction of the inelastically deformed sites is less than 1, see Eqs. (11)-(13). At $\sigma > \sigma_0^s$ a dislocation-like loop of size $l_d$ is formed. The Burgers vector is in the loop plane.

With the increase of the shear stress two types of inelastic deformation can occur in the triple joint. First, the dislocation-like loop propagates into the body of neighboring cluster when $\sigma > \sigma_c^*$, as it is shown in Fig. 9c. To estimate the value of $\sigma_c^*$ we note that the local strength of cluster body is nearly equal to the theoretical limit, $\sigma_{cl} \sim 10\bar{\mu}_{cl}$, where $\bar{\mu}_{cl}$ is the mean value of the shear modulus within the cluster body. From the other hand, the shear stress at the dislocation loop is [23]

$$\sigma_{disl} = \sigma_0^s + (l_d/2a)^{1/2}(\sigma_{cl} - \sigma_0^s) \qquad (27)$$

Equating this quantity with $\sigma_{cl}$ we obtain the estimate of $\sigma_c^*$,

$$\sigma_c^* = \sigma_0^s + (2a/l_d)^{1/2}(\sigma_{cl} - \sigma_0^s) \qquad (28)$$

Irreversible plastic deformation of the polycluster is initiated in places where the condition $\sigma > \sigma_c^*$ is fulfilled. It is seen that $\sigma_c^*$ depends on the strength of intercluster boundaries, the strength of cluster body and the cluster size, $\sigma_c^* = \sigma_c^*(\sigma_0^s, \sigma_{cl}, l)$. Substituting the mean value of the cluster size for $l$ in (28) we get a crude estimation of the macroscopic strength of the polycluster. It is worth notion that equation (28) is the Hall-Petch relation for the yield stress. In addition, initial mean DLESZ size, $l_d$, is incresing when the DLESZ is climbing into the body of neighboring cluster at $\sigma \geq \sigma_c^*$, ($l_d \geq l_{cl}$). Therefore the yield stress decreases $\sim l_d^{-1/2}$. With that an overshoot on the stress-strain curve occurs; that is, the stress decreases after its maximum value is reached [4].

The second mode of the polycluster plastic deformation is connected with the initiation of cracks at the triple joints, as shown in Fig. 9d. Cracks are formed if the Griffits condition

$$\sigma_\perp = \sigma_\perp^* \sim [2a(\varepsilon_s - \varepsilon_b)E]^{1/2} \qquad (29)$$

is fulfilled. Here

$$\sigma_\perp = \sigma_0^s + (l/2a)^{1/2}(\sigma - \sigma_0^s) \qquad (30)$$

is the stress normal to the pining boundary layer in the triple joint; $E$ is the Young modulus of the cluster; $\varepsilon_s$ and $\varepsilon_b$ are the surface energy and the boundary energy respectively.

Formation of cracks accompanies the rotational motion of a cluster. Both shear transformation and crack formation play a leading part in formation of shear bands.

Conditions (28) and (29) determine necessary conditions of the low temperature plastic deformation of MGs. They help to understand why and how the macroscopic plastic deformation starts, but we do not yet have the equation describing the macroscopic plastic deformation process (similar to those formulated by Langer et al [10-13], which describe the localization of shear transformations i.e. the shear band formation.

Let us estimate the yield stress of the polycluster, $\sigma_Y^{pc}$. The "soft" boundaries diminish the mean value of the macroscopic shear modulus of MG. As a crude estimate we can write

$$\bar{\mu} = \mu_{cl}(1 - c_{bond}) \qquad (31)$$

where $c_{bound}$ is the fraction of atoms belonging to the boundary layers. If the boundary layer width is about $2a$ and the cluster size $l_{cl} \approx 40a$ then $c \approx \sigma a / l_{cl} \sim 0.15$ and $\bar{\mu} \sim 0.85 \mu_{cl}$. With $\sigma_{cl}^s \gg \sigma_s^0$ we have $\sigma_Y^{pc} \sim 10^{-1} \bar{\mu} (2a/l)^{1/2} \sim 0.025 \bar{\mu}$. This estimate agrees with experimental evidences [4].

## 4. Discussion

In the model of dislocationless sliding formulated above we have considered an infinite sliding layer. In this form the model can be also applied to description of the frictional force of a friction couple with known surface roughness and adhesion bond. Protrusions on rough sliding surfaces are distributes in a wide range [29]. Therefore to calculate the sliding velocity, see Eq. (5), we have to find both the effective time $\tau_{sl}$ and the effective scale of elementary displacement $<d>$.

In spite of some unclear features of structure of nano-crystalline metals, their mechanical properties have much in common with MG properties. Furthermore, the plastic deformation mode of nanocrystals depends essentially on grain size $d_G$. At $d_G > d_G^c \sim 15$ nm the yield stress depends on $d_G$ in accordance with the Hall-Petch relation. However, at $d_G < d_G^c$, the yield stress becomes proportional to $d_G$ [30,31]. We assume that at small $d_G$ the deformation mode is controlled by grain boundary sliding during grain rotations. In this case boundary reconstructions play a role of lubricant redistribution that diminishes inner friction of the grain boundary creep. The detailed description of this mode will be given elsewhere.

## 5. Conclusions

- Master equations of the homogeneous sliding in the layer with a random microscopic potential relief have been exactly solved.
- The characteristic mechanical quantities have been defined implicitly in the form of functionals of the distribution function of local critical stresses.
- Mechanisms of low temperature plastic deformation of polycluster MGs initiated by intercluster boundary sliding have been described.

**Appendix 1**

Let us consider the model of a particle in the two-level potential which is characterized by two barrier parameters, $E_+$ for the direct jump and $E_-$ for the reverse one, Fig. 3. In a random potential relief the parameters $E_+$ and $E_-$ are the random variables. The transitions of the particle from a site to another site are assumed to represent a Poisson process in time and described by the kinetic equation

$$\frac{dP_2}{dt} = \nu_1^0 \exp(-\beta E_+) P_1 - \nu_2^0 \exp(-\beta E_-) P_2 \qquad (32)$$

where $P_i$ is the probability to find a particle in the site $i = 1,2$; $\nu_i^0$ is the vibration frequency, $\beta = 1/k_B T$, $T$ is the temperature and $k_B$ is the Boltzmann constant.

It is worthwhile to introduce the following parameters: $E_i = (E_- + E_+)/2$, and $\sigma_{in} \nu_a = E_+ - E_-$, where $E_i$ is the average energetic barrier and the parameter $\sigma_{in}$ can be interpreted as an internal stress; $\nu_a$ is the activation volume of elementary single jump which is assumed to be of about atomic volume.

Applied external shear stress $\sigma_e$ changes barrier heights $E_+$ and $E_-$

$$\begin{aligned} E_+ &= E_i + \nu_a(\sigma_{in} - \sigma_e)/2 \\ E_- &= E_i - \nu_a(\sigma_{in} - \sigma_e)/2 \end{aligned} \qquad (33)$$

Since $P_1 + P_2 = 1$, we can set $P_2 = P$ and $P_1 = 1 - P$. Then equation (32) subject to (33) becomes

$$\frac{dP}{dt} + \omega(t) P(t) = \Omega(t), \quad P(0) = 0 \qquad (34)$$

where

$$\omega(t) = v^0 \exp(-\beta E_i) 2\operatorname{ch}(\beta\lambda(t)), \quad \Omega(t) = v^0 \exp[-\beta E_i + \beta\lambda(t)], \quad \lambda(t) = (\sigma_e(t) - \sigma_{in})v_a/2 \quad (35)$$

When the external stress is a slow changing function of time,

$$\lambda'(t) << \Omega(t) ch^2(\lambda(t)) \qquad (36)$$

the solution of Eq. (34) reads

$$P(t) = \frac{\Omega(t)}{\omega(t)} - \frac{\Omega(0)}{\omega(0)} \exp(-\kappa(t)), \quad \kappa(t) = \int_0^t \omega(\tau) d\tau \qquad (37)$$

If the external stress $\sigma_e$ does not depend on time and

$$\sigma_e > \sigma_i = 2E_i/v_a + \sigma_{in} \qquad (38)$$

the atomic configuration goes from a site to a neighbor site without thermal activation. The parameter $\sigma_i$ is the critical shear stress at site $i$.

Combining Eqs. (4) and (35) we have

$$\omega(t) = v^0 \exp\!\left[-\alpha\!\left(\sigma_i - \sigma_e^i\right)\right] + v^0 \exp\!\left[-\alpha\!\left(\sigma_i^{(-)} + \sigma_e^i\right)\right]$$
$$\Omega(t) = v^0 \exp\!\left[-\alpha\!\left(\sigma_i - \sigma_e^i\right)\right] \qquad (39)$$

Here $\sigma_i^{(-)} = 2E_i/v_a - \sigma_{in}$. Note that parameter $\sigma_i^{(-)}$ have the same distribution function as parameter $\sigma_i$, i.e. $g(\sigma) = g(\sigma^{(-)})$.

The general solution (37) can be used for time dependent stressing e.g. reciprocating sliding. Here we assume that external stress $\sigma_e$ is time-constant and $\sigma_e/(1-\Delta) > \sigma_{in}$. These assumptions give $\Omega(t)/\omega(t) = 1$ and relation (6) for $P(t)$.